\documentclass[reprint,superscriptaddress,nofootinbib,amsmath,amssymb,aps,prl,onecolumn,notitlepage]{revtex4-1}

\usepackage[a4paper, margin=2.50cm]{geometry}

\usepackage{bbm}
\usepackage{graphicx}
\usepackage{fancyhdr, color}
\usepackage{hyperref} 

\newcommand{\chqfluct}{{\color[rgb]{0.4,0.2,0.9} \sc Qbook:Ch.9}}

\begin{document}

\fancyhead[C]{\sc \color[rgb]{0.4,0.2,0.9}{Quantum Thermodynamics book}}
\fancyhead[R]{}

\title{Quantum thermodynamics in a single-electron box}

\author{Jonne V. Koski}
\email{koskij@phys.ethz.ch} 
\affiliation{Department of Physics, ETH Zurich, CH-8093 Zurich, Switzerland.}

\author{Jukka P. Pekola}
\email{jukka.pekola@aalto.fi} 
\affiliation{QTF Centre of Excellence, Department of Applied Physics, Aalto University School of Science, P.O. Box 13500, 00076 Aalto, Finland}

\date{\today}

\begin{abstract}
This chapter provides an overview of the methods and results for quantum thermodynamic experiments with single-electron devices. The experiments with a single-electron box on Jarzynski equality and Crooks relation, two-temperature fluctuation relations, and Maxwell's demon performed over the past few years are reviewed here. We further review the first experimental realization of an autonomous Maxwell's demon using a single-electron box as the demon. 
\end{abstract}

\maketitle

\thispagestyle{fancy}

In an electronic system, the thermodynamic quantities of heat and entropy are transfered by electrons, while work is done on the system by externally applied potentials. These systems are very well defined with a simple, easily controllable Hamiltonian. The crucial ingredients to conduct the experiments are the ability to track the transitions of single electrons, or to infer the transferred heat from a change in local electronic temperature. The immediate benefit of an electronic setup is the robustness of the device, permitting hundreds of thousands of repetitions and therefore reliable statistics of the chosen thermodynamic process in contrast to experiments with molecular or colloidal particles, where the number of repetitions is typically limited to hundreds. 

The chapter is outlined as follows. 
The first four sections give the principles and methods required to design and execute thermodynamic experiments with a single-electron box. The last four sections review the recent thermodynamic experiments conducted in that setting.

\section{Heat in electronic systems} \label{sec:Heat}

In this chapter we are mainly concerned with systems where finite heat baths are formed of electrons in an island of normal metal \cite{rmp}. In practise these reservoirs are thin film metals fabricated by electron-beam lithography. Due to manufacturing constraints, the islands have typically a volume of $10^{-3}$ $\mu$m$^3$ or larger, which means that there are of the order of or more than $10^9$ conduction electrons in this system. Thus it is fair to say that such an island serves as a body to which temperature can be assigned, at least in equilibrium. Thermal electrons in this dot are only weakly coupled to the rest of the circuit. Typically this is made possible by embedding it in a low temperature ($\ll 1$ K) environment, where coupling to phonons, which typically scales with temperature as $T^5$ \cite{wellstood}, is extremely weak. 
For many experiments to be presented below, namely the ones based on electron counting, this weakness of the coupling is not essential though. In all the experiments and their analysis, it is, however, important that the relevant relaxation timescales have a certain hierarchy. In particular, it is essential that the electron-electron (e-e) relaxation time is the fastest one in the system. For standard metallic structures, this relaxation time is $10^{-9}$ s or shorter \cite{pothier}, which makes it about four orders of magnitude faster at the said temperatures than the electron-phonon (e-p) relaxation time \cite{wellstood}, the other fundamental relaxation process in this system. If furthermore the external drive of the system is slow in comparison to the e-e relaxation rate, one may always assume that the electrons in the “absorber” normal metal form a system with well-defined local temperature at all instants of time.

\begin{figure} [h!]
\includegraphics[width=12cm]{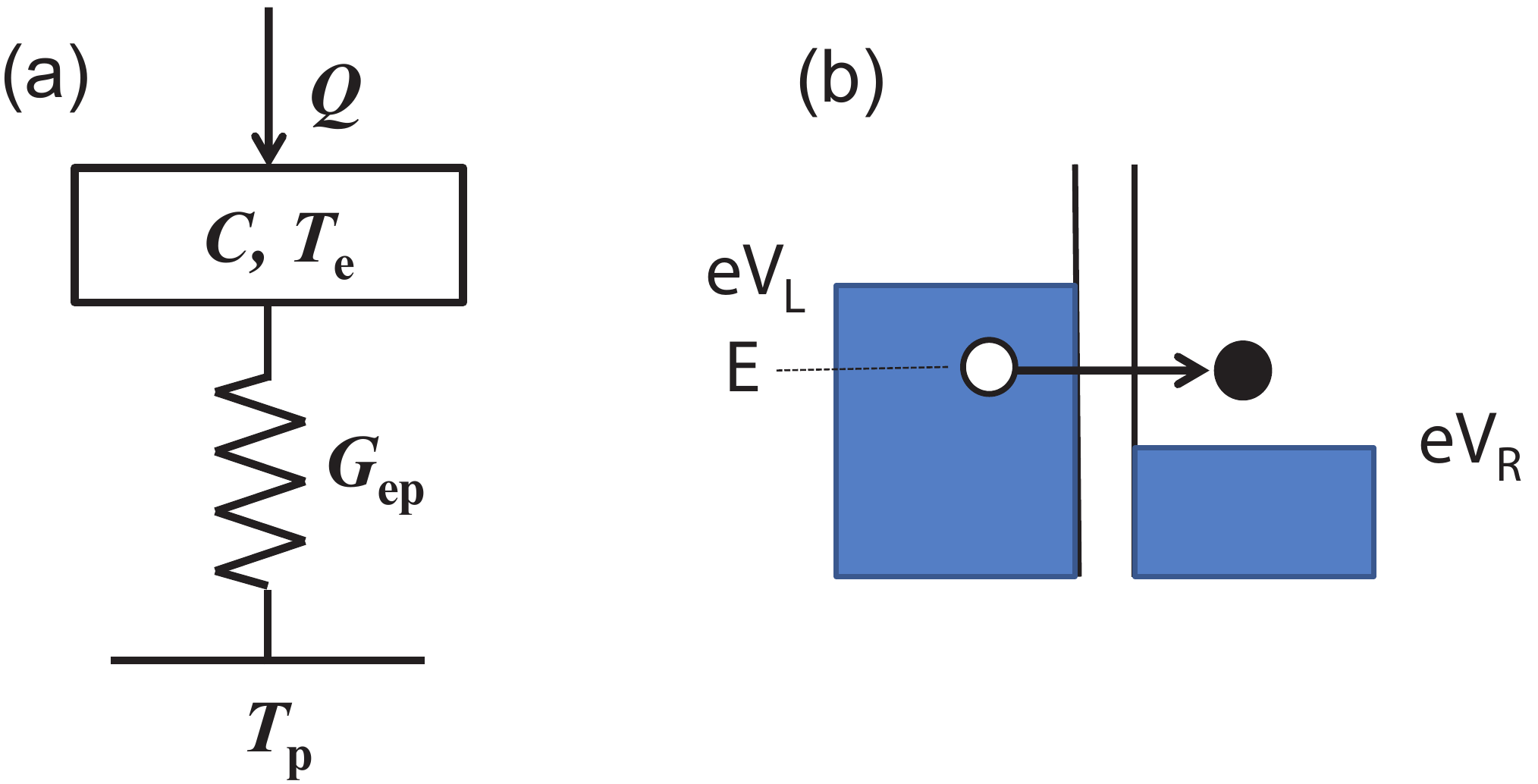}
  \caption{Schematics of the systems discussed. (a) Typically a finite electron system (metal conduction electrons) with heat capacity $C$ and at temperature $T_\text{e}$ interacts weakly with the phonon bath at temperature $T_{\rm p}$ via thermal conductance $G_{\rm ep}$. Electrons themselves interact to establish internal thermal equilibrium. The electron system is subject to various heat inputs $Q$. (b) A tunnel barrier with chemical potential difference $eV$, where $V=V_L-V_R$. An electron tunnels at energy $E$, creating an excitation (heating) in both electrodes.}
 \label{fig1}
\end{figure}

\section{Charge and heat transport across a tunnel junction} \label{sec:Junction}

In stochastic thermodynamics experiments performed up to now, heat has been measured by indirect means \cite{jp}. In other words, one typically relies on measurement of charges, voltages or currents in electrical circuits or on positions or momenta in mechanical systems, which allow one to evaluate the heat via a model applicable to the system in question. Direct measurement of stochastic heat is still elusive, although it looks feasible at least in a low temperature experiment on electric circuits as presented here \cite{njp}. Single-electron systems, in particular a single-electron box (SEB), see next chapter, provide a well characterized set-up to investigate stochastic thermodynamics either by indirect means but also by direct measurement of heat in the future. 

A basic element in (single-)electronic circuits is a tunnel barrier, which in our metallic systems separates two Fermi seas of electrons, see Fig. 1. A chemical potential difference $\Delta\mu =eV$ given by the voltage difference $V=V_L-V_R$ is applied across the barrier between reservoirs $L$ and $R$. Under these conditions, a tunneling event through the barrier leads to dissipation. If the electron tunneling from the left lead has energy $E$ with respect to the Fermi level at $eV_L$, the energy deposited to this reservoir equals $Q_L=eV_L-E$. Now assuming an elastic process, with no exchange of energy by the electron in the process (horizontal arrow in Fig. 1), the electron lands on the right reservoir at the energy $E-eV_R$, which is also the energy input $Q_R$ to this lead. Although the energies deposited to $L$ and $R$ vary stochastically depending on the energy $E$ of the tunneling electron, the total heat input to the system formed of the left and right sides of the junction is constant equal to $Q = Q_L+Q_R = e(V_L-V_R) = eV$, as one would naively expect. In terms of the total heat $Q$, the stochasticity of the process is then determined by the number of electrons tunneling during the observation time \cite{kung}. Alternatively, under time-dependent driving protocols, it is determined by the instantaneous value of $V(t)$ at the time instant when the electron tunnels \cite{saira}.

The rate of tunneling from $L$ to $R$ is given by a golden- rule based expression
\begin{equation} 
\label{rate}
\Gamma =\frac{1}{e^2R_T}\int_{-\infty}^\infty dE n_L(E-eV_L)n_R(E-eV_R)f_L(E-eV_L)[1-f_R(E-eV_R)],
\end{equation}
where $R_T$ is the tunnel resistance of the barrier (determined by the properties of the junction), $n_{i}(E)$ is the (normalized) density of states of electrons, and $f_{i}(E)$ the distribution of electrons in lead $i=L,R$. The two types of conductors considered in this chapter are normal metal (N, copper), for which $n_\text{N}(E) = 1$, and a superconductor (S, aluminum), for which $n_\text{S}(E) = \text{Re}(|E| / \sqrt{(E)^2  - \Delta^2})$, where $\Delta$ is the superconductor energy gap \cite{BCS}. The main feature of $n_\text{S}(E)$ is that the density is zero for the energy range $-\Delta \leq E \leq \Delta$. If electrons in each lead are internally in equilibrium, they form a Fermi-Dirac distribution $f_i(E)=1/(1+e^{E / k_BT_i})$, where the temperatures of the two leads are given by $T_L$ and $T_R$, respectively. Tunneling given by Eq. \eqref{rate} is stochastic, and in the case of no correlations and assuming a fixed chemical potential difference $eV$ across the barrier, it is a Poisson process. 

For a more general description, it is convenient to view the transition rate for a specific energy cost $\Delta E$, that is, the difference in energy between the final and initial state. In the case of a voltage $V$ biased tunnel junction where the electron is initially at potential $V_L$ and finally at potential $V_R$, we have $\Delta E = eV_R - eV_L$. This translates the tunneling rate by Eq. \eqref{rate} to
\begin{equation} 
\label{energyrate}
\Gamma_{L\to R}(\Delta E) =\frac{1}{e^2R_T}\int_{-\infty}^\infty dE n_L(E)n_R(E-\Delta E)f_L(E)[1-f_R(E-\Delta E)].
\end{equation}
In accordance to thermodynamic principles, when the two leads are at equal temperatures $T_L = T_R = T$, the tunneling rate obeys detailed balance $\Gamma_{L\to R}(\Delta E) / \Gamma_{R \to L}(-\Delta E) = \exp(-\Delta E / k_\text{B} T)$.
 
\section{Single-electron effects} \label{sec:SEeffects}

As covered in the previous section, heat is primarily transported in nanoelectronic circuits by electrons. Those electrons are subject to mutual Coulomb interaction that can influence the properties of the device dynamics. As discussed below, when the size scale and the temperature of the circuit is small, this interaction becomes the dominant effect and the total electron number in the system is critical at the precision of one electron. Such devices are in general known as single-electron devices, of which the most relevant for this chapter are the single-electron box and the single-electron transistor.


A single-electron box (SEB) consists of a single piece of metal - an 'island' - connected to a grounded metallic lead by a tunnel junction and coupled to a gate electrode with capacitance $C_g$. The tunnel junctions allows electrons to tunnel into and out of the island, changing the total charge of the island by $\pm e$. When an electron tunnels to the island, further electrons that would follow are repelled by the previously added negative charge. This effect is characterized by charging energy $E_\text{ch} = \frac{(-en)^2}{2 C} \equiv E_C n^2$, where the net charge on the island is $-en$ for $n$ electrons, and $C$ is the total capacitance of the island. The charging energy for a single-electron, $E_C = e^2 / 2C$, is the characteristic energy scale of the SEB. The island charge can be manipulated by gating, i.e. by tuning the gate electrode potential $V_g$. If the applied potential is positive, it will attract a negative charge equal to $-C_g V_g$ to the island. This corresponds to an effective gate number $n_g = C_g V_g / e$ that can be non-integer as it describes charge rearrangement on the island adapting to the surrounding potential rather than actual number of electrons. The Hamiltonian of the SEB is then
\begin{equation} H = E_C (n - n_g)^2. \label{eq:ChargingEnergy} \end{equation}


An SEB constitutes a controllable two-level system as follows. The system degree of freedom is $n$ and the control parameter is $n_g$.  If the charging energy is substantially larger than thermal energy, $E_C \gg k_\text{B}T$, only the lowest energy states need to be considered. Note that the charging energy by Eq. \eqref{eq:ChargingEnergy} remains constant if both $n$ and $n_g$ are offset by an integer. We can therefore consider $n_g$ to be operated in the range $0...1$, such that the single-electron box forms a two-level system with possible states $n = 0$ or $n = 1$. The energy difference between the two states can be readily controlled with $V_g$ and is given by Eq. \eqref{eq:ChargingEnergy} as $\Delta E = H_{n = 1} - H_{n = 0} = - 2E_C (n_g - 0.5)$. $\Delta E$ directly determines the transition rate for the event $n:0\to 1$ (an electron enters the island, changing the number of electrons from 0 to 1) as $\Gamma_{L\to R}(\Delta E) \equiv \Gamma_{0 \to 1}$, and for $n:1\to 0$ as $\Gamma_{R\to L}(-\Delta E) \equiv \Gamma_{1 \to 0}$ from Eq. \eqref{energyrate}. The control parameter value $n_g = 0.5$ sets the two states to have equal energy, and is often referred to as the degeneracy point. As thermal fluctuations are relevant when $|\Delta E| \lesssim k_\text{B}T$, most thermodynamic processes take place around this point.


Next we consider the validity of the two-level condition, i.e. the criterion $E_C \gg k_\text{B} T$, for practical nanoscale devices. If the SEB island with a maximum length scale of $l = 1~\mu$m lies on a silicon/silicon oxide substrate (self capacitance $C_0 \sim (\varepsilon_\text{Si} + \varepsilon_0)  l / 2 \approx 50$ aF) and is tunnel coupled to a metallic lead through an aluminum oxide insulator layer with a thickness of $d = 2$ nm and a cross section area of $A = 100 \times 100$ nm$^2$ (capacitance $C_J = \varepsilon_{\text{Al}_2\text{O}_3} A/d \approx$ 440 aF), the single electron charging energy is $E_C = e^2 / 2(C_0 + C_J) \approx k_\text{B} \times 2$ K. Dilution cryostats can routinely reach temperatures of $T \ll 50$ mK well ascertaining the condition $E_C \gg k_\text{B} T$. A standard SEB therefore realizes a two-level system as the probability to be in a state other than $n = 0$ or $n = 1$ is $P < 10^{-17}$.

\begin{figure} [h!]
\includegraphics[width=12cm]{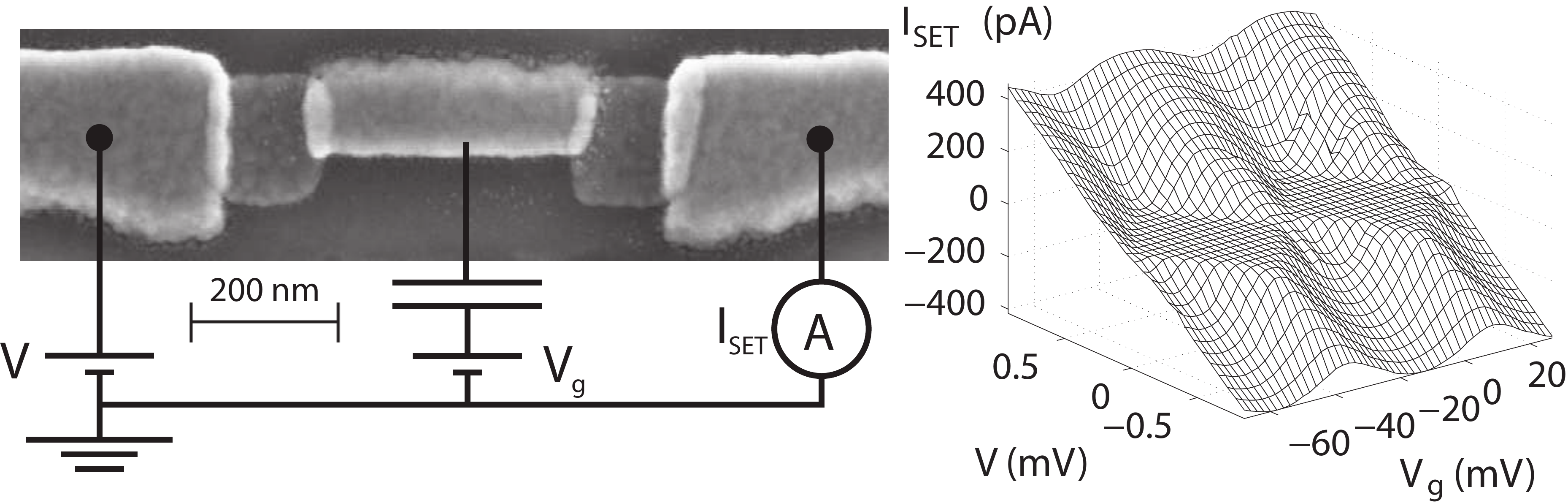}
\caption{Left panel: An example of a single-electron transistor. A metallic island (structure in the middle) is connected to two metallic leads by tunnel junctions. The SET is voltage $V$ biased, and the current $I_\text{SET}$ is manipulated by the gate voltage $V_g$. Right panel: Measured current through the SET as a function of $V$ and $V_g$. Reproduced from \cite{jvkapl}, with permission from AIP Publishing. }
 \label{fig:SET}
\end{figure}


A single-electron transistor (SET) is similar to an SEB, with the distinction that the island is connected to two (or more) metallic leads. An example device is shown in Fig. \ref{fig:SET}. The two leads can further have potentials $V_\text{L} = V / 2$ and $V_\text{R} = -V / 2$, where L(R) refers to the left (right) lead and $V$ is the potential bias. The charge current from the left to the right lead occurs under a finite bias $V$ and is controlled with the gate voltage $V_g$, as seen in Fig. \ref{fig:SET}. When $|\Delta E| > eV / 2$, charge transport is suppressed as Coulomb blockade prevents electrons from tunneling into (out of) the island, while with $|\Delta E| < eV /2$, electrons can consecutively tunnel in from the left lead, and out to the right lead. In general, the full dynamics can be modeled with a rate equation,
\begin{equation} 
d P(n) / dt = \Gamma^\text{out}_{n + 1 \to n}P(n + 1) + \Gamma^\text{in}_{n - 1 \to n} P(n - 1) - \left(\Gamma^\text{out}_{n \to n - 1} + \Gamma^\text{in}_{n \to n + 1}\right) P_n,
\end{equation}
where $P_n$ is the probability to occupy state $n$, and $\Gamma^\text{in (out)}_{n \to n \pm 1}$ is the transition rate in (out) from the island, changing the state from $n$ to $n \pm 1$. The probability distribution $P_n$ can be solved under steady state $d P(n) / dt = 0$ and normalization $\sum P_n = 1$. The current to lead R is then given by
\begin{equation} I_\text{R} = \sum_n \Gamma^\text{out, R}_{n \to n + 1}P(n + 1) -  \Gamma^\text{in, R}_{n + 1 \to n}P(n), \end{equation}
where the superscript R refers to transitions occurring by tunneling events through the right junction. Note that at steady state, $I_\text{L} = -I_\text{R}$.


\section{Detecting the charge in a single-electron box}

\begin{figure} [h!]
\includegraphics[width=4.5cm]{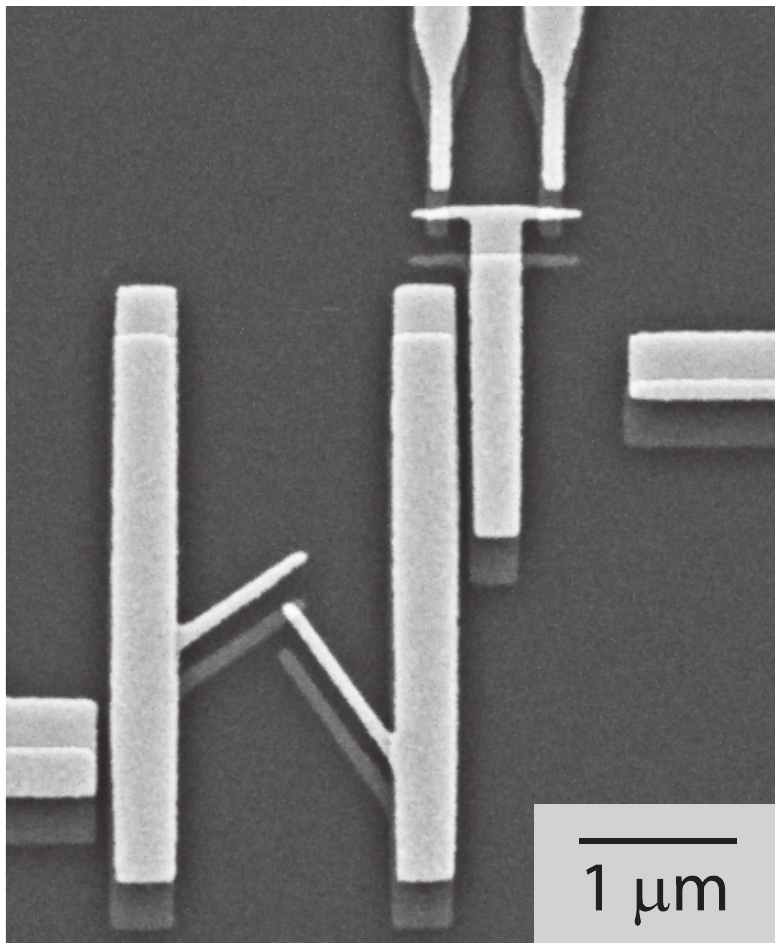}~~~ 
\includegraphics[width=7.5cm]{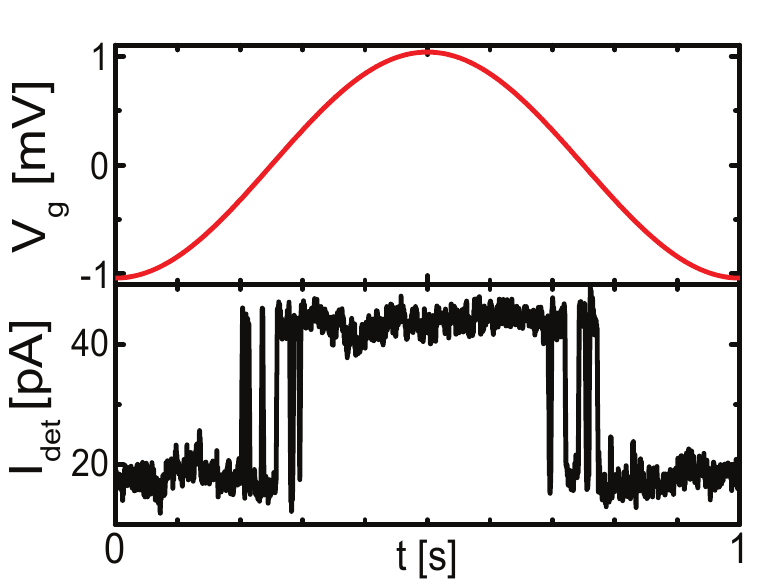}
  \caption{Left panel: A scanning electron micrograph of a single-electron box coupled to a single-electron transistor. The light gray features are copper, and the gray 'shadows' of the same features are aluminum covered by an insulating aluminum oxide layer. The H-shaped SEB has a charging energy of $E_C \approx 2$ K and is cooled down to 200 mK and below, ensuring that it behaves as a two-level system. A tunnel junction is connecting the two halves of the "H". The SET, which is the fork-shaped structure on the top, lies close to the SEB, resulting in capacitive coupling that facilitates charge detection. Right panel: modulating the gate voltage $V_g$ triggers an evolution of the charge state $n(t)$ observed as jumps in the detector current $I_\text{det}$. Reproduced figure with permission from \cite{saira}. Copyright 2012 by the American Physical Society.}
 \label{fig:JEBox}
\end{figure}

An SEB provides a controllable two-level system, however an ability to time-dependently resolve its charge state is essential for the execution of thermodynamic experiments. 
This is realized with a charge detector in form of an SET by coupling the islands of the two devices with capacitance $C_\text{int}$, as illustrated in Fig. \ref{fig:JEBox}. 
The two coupled devices follow a Hamiltonian
\begin{equation} H = E_C(n - n_g)^2 + E_C^\text{det}(n^\text{det} - n_g^\text{det})^2 + J(n - n_g)(n^\text{det} - n_g^\text{det}). \label{eq:coupled_devices} \end{equation}
Here, the superscript 'det' refers to the SET, and $J = e^2 C_\text{int} / (C C_\text{det} - C_\text{int}^2)$ describes the mutual Coulomb interaction between the electrons in the two islands. 
An electron tunneling in the SEB ($\Delta n = \pm1$) effectively acts as a gate voltage change $\Delta n_{g, \text{eff}}^\text{det} = -J \Delta n / (2 E_C^\text{det})$ on the SET. 
Therefore, when the SET is voltage $V$ biased and operated in a regime of finite current flow, its current switches between two values $I_0$ and $I_1$ that correspond to the two states $n = 0$ or $n = 1$ whenever an electron tunnels in the SEB. 
By measuring a time trace of the current through the SET during a thermodynamic process, the evolution of the system state $n$ is determined in real time as demonstrated in Fig. \ref{fig:JEBox}. 


Practical charge detection schemes are limited by the bandwith of the detector. This is typically set by the cut-off frequency of a low-pass filter, either determined by the measurement setup, or by noise that has to be averaged out to reach a sufficient signal-to-noise ratio for exact determination of the charge state. For a standard SET charge detector, a typical bandwidth is up to 1 kHz, and the frequency of tunneling events in the system should be lower for an accurate time-resolved determination of the system state evolution. A typical tunnel junction resistance is of the order of 100 k$\Omega$, which means that a fully normal metallic SEB would not qualify for these experiments. At 50 mK temperature, a normal metallic SEB would give a tunnel rate \eqref{rate} of $300$ MHz at the degeneracy point, significantly exceeding the detection bandwidth. However, by fabricating either the SEB island or the lead (or, in case of a two-island SEB, one of the islands) out of aluminum (superconducting below 1 K), the superconductor energy gap $\Delta \simeq$ 200 $\mu$eV drastically suppresses the tunneling rates to 1 - 1000 Hz level around the degeneracy point, bringing the events to a detectable range. This method is used in the experiments reviewed in the next three sections.


With the charge detection scheme discussed above, the system state as a function of time, $n(t)$, can be resolved. Other thermodynamically relevant quantities, namely the state occupation probability and transition rates for a given control parameter $n_g$, are determined from long time traces of $n(t)$ for a stationary $n_g$. The trace duration $\mathcal{T}$ should cover a sufficient number of tunneling events to gain reasonable statistical reliability. Then the steady state occupation probability $P_n(n_g)$ is determined from the total time $\mathcal{T}_n$ spent on the state $n$ as $P_n(n_g) = \mathcal{T}_n / \mathcal{T}$.


The tunneling rates $\Gamma_{0 \to 1}$ and $\Gamma_{1 \to 0}$ are extracted by determining the lifetime of states $0$ and $1$, that is, the average time that the system spends in state $0$ or $1$ before transitioning to the other state. First, the time instants and the initial and final states of tunneling events are extracted from a time trace with stationary $n_g$. For each recorded tunneling event, the time spent on the state $n$ to which the event brought the system is determined as the time $\tau_n$ until the next tunneling event occurs. The lifetime of state $n$ is the average $\langle \tau_n \rangle$. The corresponding tunneling rate is then determined as $\Gamma_{n \to (1 - n)} = 1 / \langle \tau_n \rangle$.
For a given $\Delta E$ as determined by the control parameter $n_g$ and at uniform electronic temperature $T_\text{e}$, the tunneling rates follow detailed balance, $\ln\left(\Gamma_{0 \to 1} / \Gamma_{1 \to 0}\right) = -\Delta E / k_\text{B} T_\text{e}$. Assuming that the SEB electrodes are well thermalized to the cryostat temperature so that $T_\text{e}$ is known, the calibration for $\Delta E(V_g)$ is obtained from the detailed balance condition.


\section{Single-electron box as a testbed for quantum thermodynamics}


This section gives a brief overview of the experimental test in \cite{saira} on Jarzynski equality \cite{jarzynski} (JE), $\langle \exp(-W / k_\text{B}T) \rangle = \exp(-\Delta F / k_\text{B} T)$, where $W$ is the applied work and $\Delta F$ is the free energy difference, and Crooks relation \cite{crooks}, $\ln(P(W) / P_R(-W)) = (W - \Delta F) / k_\text{B} T$, where $P(W)$ is the probability to apply work $W$ in the forward process and $P_R(-W)$ is the probability for negative work in the reverse process. The test is carried out by introducing a drive protocol to the SEB that transfers an electron from the left island to the right, and a reverse protocol that does the opposite. Each realization of the protocol requires work and generates heat stochastically. The statistics of those quantities are studied by performing multiple realizations of the drive protocol.


More specifically, the SEB shown in Fig. \ref{fig:JEBox}, is driven by modulating the control parameter as $n_g(t) = 1 / 2 - \cos(2 \pi f t) / 2$ while continuously monitoring the charge state $n(t)$ with a nearby SET.
This drive is symmetric around the degeneracy point $n_g=1/2$.
A single realization of the drive protocol is taken from $t = 0$ to $t = 1 / (2f)$, during which
the system is driven across the degeneracy point from $\Delta E = E_C$ where $n = 0$ is the ground state to $\Delta E = -E_C$ where $n = 1$ is the ground state. 
The reverse process takes place from $t = 1 / (2f)$ to $t = 1 / f$ bringing the state from $n = 1$ to $n = 0$. 
Thus each period of drive consists of two identical realizations of the protocol.
One of the initial criteria to test JE is to start from thermal equilibrium, which is the case in the experiment since the system always starts from the ground state under the given experimental conditions. 

\begin{figure} [h!]
\includegraphics[width=10cm]{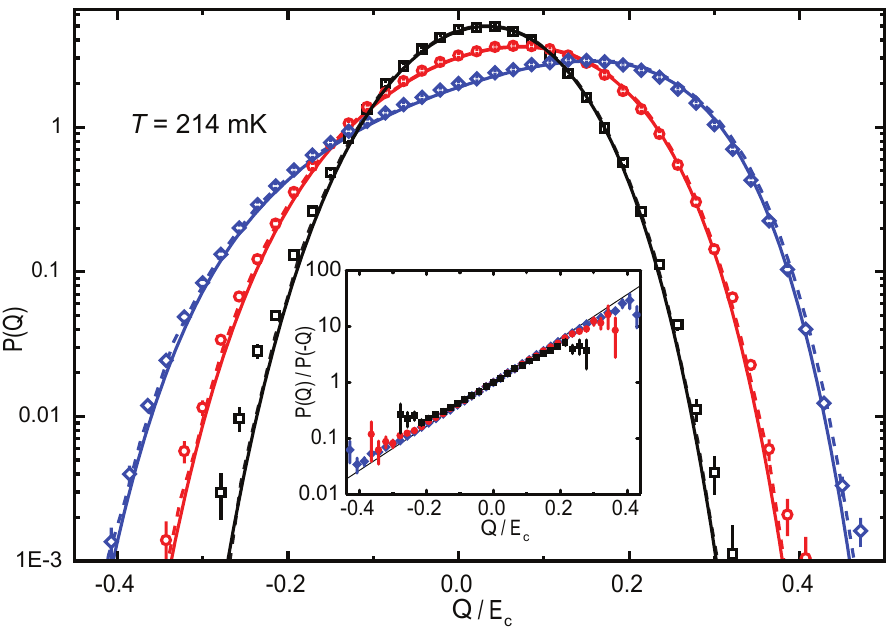}
\caption{Measured probability distributions for dissipated heat, i.e. histograms of measured $Q$ for drive frequencies $f = 1$ Hz (black, squares), 2 Hz (red, circles), and 4 Hz (blue, diamonds). Symbols show measured data, solid lines give theoretical predictions, and dashed lines include a finite detector bandwidth in the model. Inset: The test of Crooks relation. Reproduced figure with permission from \cite{saira}. Copyright 2012 by the American Physical Society.}
 \label{fig:JEDistributions}
\end{figure}


For each realization of the drive protocol the relevant thermodynamic quantities, work and heat, are determined from the measured trace $n(t)$ as follows. 
As mentioned in the second section, heat is generated at every electron tunneling event that takes place during the process as an electron carries an energy $E$ from source to target electrode, and the excess energy is distributed to the reservoirs as heat.
Correspondingly, in a tunneling event, the energy of the system changes by $\Delta E$ as determined by the control parameter $n_g$. 
Energy conservation requires that the energy added to the system must originate from the heat baths, generating heat $Q = -\Delta E(n_{g})$.
If a single trace has multiple tunneling events, indexed with $i$, the total heat generated is
\begin{equation} Q = -\sum_i \Delta n_{i} \Delta E(n_{g, i}), \label{eq:heat} \end{equation}
where $\Delta n_{i}$ is the electron number change in tunneling event $i$ ($+1$ for $n:0\to 1$ and $-1$ for $n:1\to 0$), and $n_{g, i}$ is the value of the control parameter at the given time instant of the tunneling event. Work, in contrast, is applied when the drive changes the system energy $H$ by Eq. \eqref{eq:ChargingEnergy}. For a single trace of duration $T$, work is evaluated as
\begin{equation} W = \int_{0}^T 2 E_C(n_g(t) - n(t)) \frac{dn_g(t)}{dt}dt. \label{eq:work} \end{equation}
We note that since the processes practically always start and end in fixed states and the free energy change in the process is zero, work and heat are equal.


The experiments were carried out at various driving frequencies and at a few temperatures, and under each experimental condition more than $10^5$ repetitions of the protocol were realized. The probability distributions for $Q$ as extracted by Eq. \eqref{eq:heat} from the ensemble of traces are shown in Fig. \ref{fig:JEDistributions}. It was found that the Jarzynski equality is valid within 3\% and that the distributions obey Crooks relation. The main uncertainty stems from the finite banwidth of the detector leading to unregistered transitions.

\section{Test of fluctuation relations for two heat baths}

\begin{figure} [h!]
\includegraphics[width=11cm]{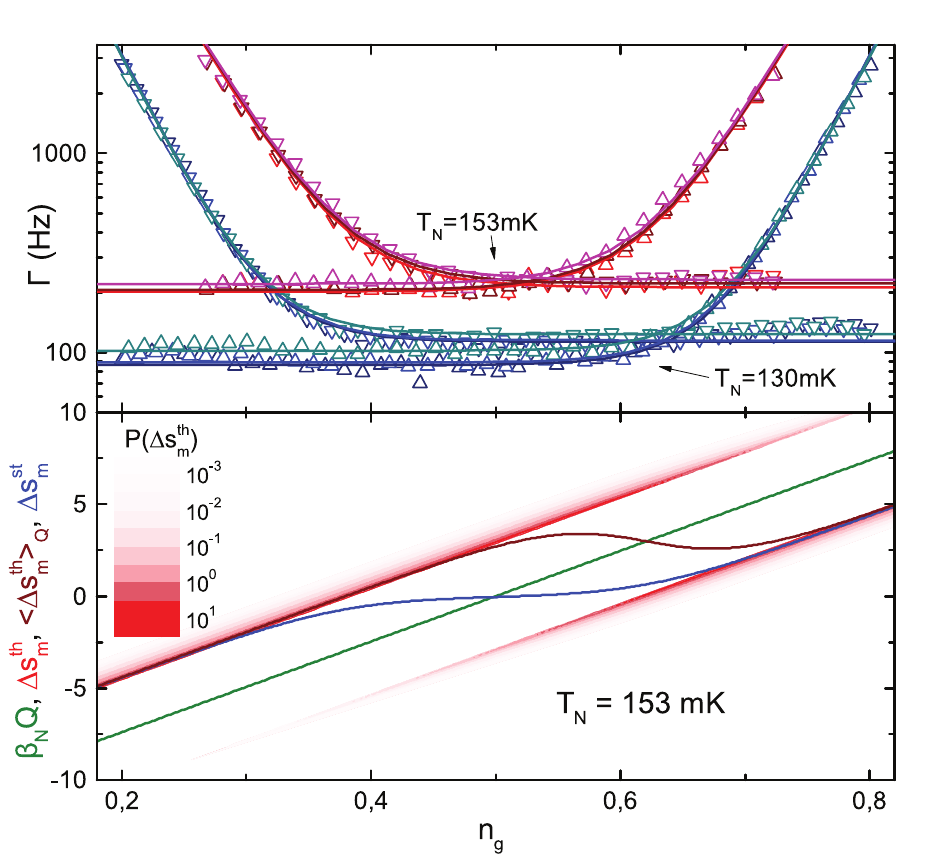}
\caption{Top panel: Measured tunneling rates for $n:0\to 1$ transitions (triangles pointing up) and $n:1\to 0$ transitions (triangles pointing down) as functions of $n_g$ at two bath tempratures $T_N$. Solid lines show corresponding fits of Eq. \eqref{rate}, yielding a superconductor temperature $T_S = 175$ mK  when $T_N = 130$ mK, and $T_S = 185$ mK when $T_N = 153$ mK. Bottom panel: thermodynamic or stochastic entropy produced in a single $n:0\to 1$ tunneling event as a function of $n_g$ at the time instant of the event. The thermodynamic entropy $\Delta s_\text{m}^\text{th}$ given by Eqs. \eqref{eq:thentropy} follows a probability distribution shown in red along the y-axis (in logarithmic scale) and is determined by Eq. \eqref{eq:energydistribution}. The average value $\langle\Delta s_\text{m}^\text{th}\rangle$ is shown in brown. The corresponding stochastic entropy production $\Delta s_\text{m}^\text{st}$ determined by Eq. \eqref{eq:stentropy} is shown in blue. For reference, the entropy $\Delta s_\text{m} = -\Delta E / k_\text{B} T_N$ that would be produced if the whole system would be at temperature $T_N$ is shown in green. For both panels, the superconductor energy gap is $\Delta \simeq 224~\mu$eV, the charging energy is $E_C \simeq162~\mu$eV, and the tunneling resistance is $R_\text{T} \simeq 1.3$ M$\Omega$. Reproduced figure from \cite{jvkNP}.}
\label{fig:TwoBathRates}
\end{figure}


The temperature of a well thermalized system is uniform. However, with an SEB, it is relatively straightforward to form a system with two unequal temperatures by maintaining a temperature difference between the two electrodes. This system allows investigation of fluctuation relations in the presence of multiple heat baths. Jarzynski equality is defined for single temperature systems and thus is not expected to hold, however, fluctuation relations for entropy production can be tested also in this situation \cite{Seifert, crooks, Schuler, Tietz}.


For the experiment \cite{jvkNP} described here, a temperature difference between the two electrodes is achieved by suppressing the thermalization mechanism of the aluminum electrode. Unlike in the SEB shown in Fig. \ref{fig:JEBox}, here the aluminum island is not covered by a normal metal shadow that would otherwise thermalize it. The only remaining channel is electron-phonon relaxation that takes place in a superconductor only via the unpaired quasiparticles. It is typical for a superconducting aluminum island at a temperature of 150 mK to host only of the order of ten unpaired quasiparticles \cite{Maisi} and it has a very weak phonon thermalization in contrast to normal metallic electrodes. The high temperature of the superconductor is verified from the measured transition rates $\Gamma_{0 \to 1}$ and $\Gamma_{1 \to 0}$, shown in Fig. \ref{fig:TwoBathRates}. By changing the cryostat bath temperature and assuming that it is the same as that of the normal metal electrode, the fits to the transition rates by Eq. \eqref{rate} indicate that the superconductor is consistently at a temperature of approximately 180 mK.

\begin{figure} [h!]
\includegraphics[width=10cm]{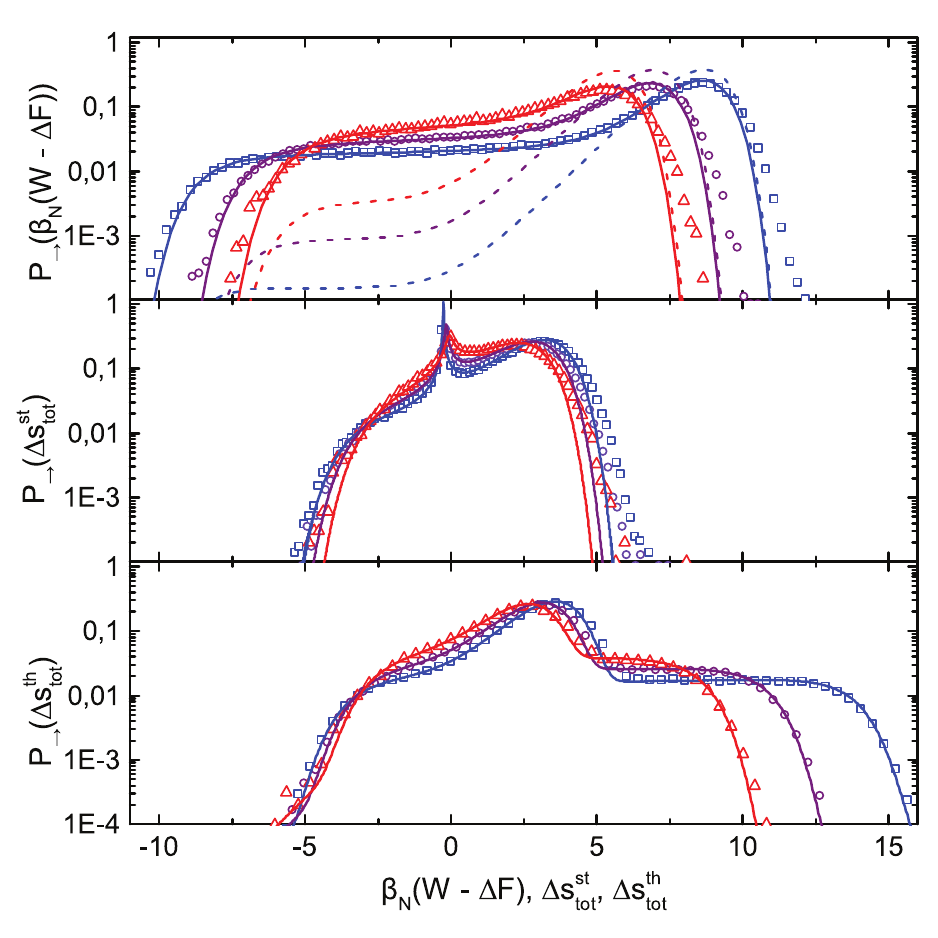}
\caption{Top panel: Measured probability distribution for $W / k_\text{B}T_N$. 
Middle panel: measured distribution for stochastic entropy production given by Eq. \eqref{eq:stentropy}. Bottom panel: distribution for thermodynamic entropy production given by Eqs. \eqref{eq:thentropy} and \eqref{eq:energydistribution} for the recorded single-jump trajectories. In all panels, data measured at $T_N = 130$ mK are shown in blue (squares), $T_N = 142$ mK in purple (circles), and $T_N = 153$ mK in red (triangles). Symbols show measured values, solid lines show theoretical predictions, and in the top panel, dashed lines show the predicted distributions for $T_S = T_N$. Reproduced figure from \cite{jvkNP}.}
\label{fig:TwoBathDistributions}
\end{figure}


For the present experiment, the drive protocol is identical to the one in the previous section. However, the feature of two unequal temperatures prompts to revise the thermodynamic quantities extracted from the measurement. First, we consider what we call 'thermodynamic' (dimensionless) entropy generated in the medium, namely
\begin{equation} \Delta s_\text{m}^\text{th} = Q_N / (k_\text{B} T_N) + Q_S / (k_\text{B} T_S), \label{eq:thentropy} \end{equation}
where $Q_N + Q_S = -\Delta E$. For unequal $T_N$ and $T_S$, the magnitudes of both $Q_N$ and $Q_S$ are relevant. As our detection scheme only allows to determine $\Delta E$ directly, we assign a probability density for heat generated in the normal electrode. For example, if we assume that an electron tunnels from the normal (N) to the superconducting (S) electrode, the probability distribution is given by
\begin{equation} P (Q_N = -\epsilon~|~\Delta E) = \frac{f_\text{N}(\epsilon) n_\text{S} (\epsilon - \Delta E) (1 - f_\text{S}(\epsilon - \Delta E))}{\int d\epsilon  f_\text{N}(\epsilon)n_\text{S} (\epsilon - \Delta E) (1 - f_\text{S}(\epsilon - \Delta E))}. \label{eq:energydistribution} \end{equation}
The resulting probability distribution for thermodynamic entropy production as a function of $n_g$ at a time instant of a tunneling event is shown in Fig. \ref{fig:TwoBathDistributions}. The distribution shows two distinct peaks for each $n_g$. Qualitatively, this can be understood from the energy gap $\Delta$ in the superconductor density of states which dictates that electrons tunneling into or out of the superconductor can either satisfy $Q_S \gtrsim \Delta$ or $Q_S \lesssim -\Delta$. 

Next, we consider what we call 'stochastic' dimensionless entropy defined in \cite{Seifert}, which for a single transition is given by
\begin{equation} \Delta  s_\text{m}^\text{st} = \ln\left(\frac{\Gamma_{0 \to 1}(n_g)}{\Gamma_{1 \to 0}(n_g)}\right). \label{eq:stentropy} \end{equation}
As seen in Fig. \ref{fig:TwoBathDistributions}, the stochastic entropy has the same value irrespective of the energy of the tunneling electron. As such, one can view this entropy as the one observed where the electron energy degree of freedom is coarse-grained. The two definitions satisfy
\begin{equation} \left\langle e^{-\Delta s_\text{m}^\text{th}} \right\rangle_{Q_\text{N}} = e^{-\Delta s_\text{m}^\text{st}}, \label{eq:entropyrelation} \end{equation}
where the $\langle ... \rangle_{Q_\text{N}}$ denotes averaging over all possible $Q_N$ with the constraint that $\Delta E$ is determined by $n_g$, and $Q_S = -\Delta E - Q_N$. From Eq. \eqref{eq:entropyrelation} it immediately follows that $\langle \Delta s_\text{m}^\text{th} \rangle \geq \Delta s_\text{m}^\text{st}$, which is also captured in Fig. \ref{fig:TwoBathDistributions}.
The total entropy produced over a trajectory is the sum over those produced in the transitions during a process, similarly as presented for the heat $Q$ in the previous section. For driven processes where both the initial and the final state follows thermal equilibrium, both definitions of entropy production are expected to follow fluctuation relations \cite{Seifert}. 


Figure \ref{fig:TwoBathDistributions} shows the measured distributions of stochastic entropy, thermodynamic entropy for single-jump trajectories, and a distribution that 'tests' JE by assuming that both electrodes are at temperature $T_N$. As expected, the JE does not hold in a system with two unequal temperatures, for the exponential average of the distribution yields  $\langle \exp(-W / k_\text{B} T_{N / S})\rangle \geq 10$. However, we find that both the stochastic and thermodynamic entropy distributions do satisfy the fluctuation relations, as the exponential average for both $\langle \exp(-\sum \Delta s^\text{m}_\text{th})\rangle \simeq 1$ and $\langle \exp(-\sum \Delta s^\text{m}_\text{st})\rangle \simeq 1$.

\section{Single-electron box in information thermodynamics}


The ability to track individual electron tunneling events in a SEB gives access to information thermodynamic experiments with processes under feedback-control. A feedback-control protocol  consists of a sequence of measurements, which determine the instantaneous system state, followed by a routine which depends on the previously measured state but is otherwise pre-determined \cite{sagawaueda, toyabe}. We consider in \cite{jvkPNAS} a simple protocol consisting of a single measurement followed by a feedback drive. The measurement obtains one bit of information from the system ($n = 0$ or $n = 1$) which, according to the thought experiment by Szilard \cite{Szilard} and Landauer's principle \cite{Landauer}, can be converted to a fundamental amount of $k_\text{B} T \ln(2)$ energy. Here, $k_\text{B}\ln(2)$ is the maximum amount of entropy for one bit of information corresponding to a case where both states have an equal probability. Indeed, Landauer's erasure principle has been demonstrated with a colloidal particle in a two-well potential \cite{Berut, Jun} by showing that erasing one bit of information expends a minimum amount of work $k_\text{B} T \ln(2)$.

\begin{figure} [h!]
\includegraphics[width=9cm]{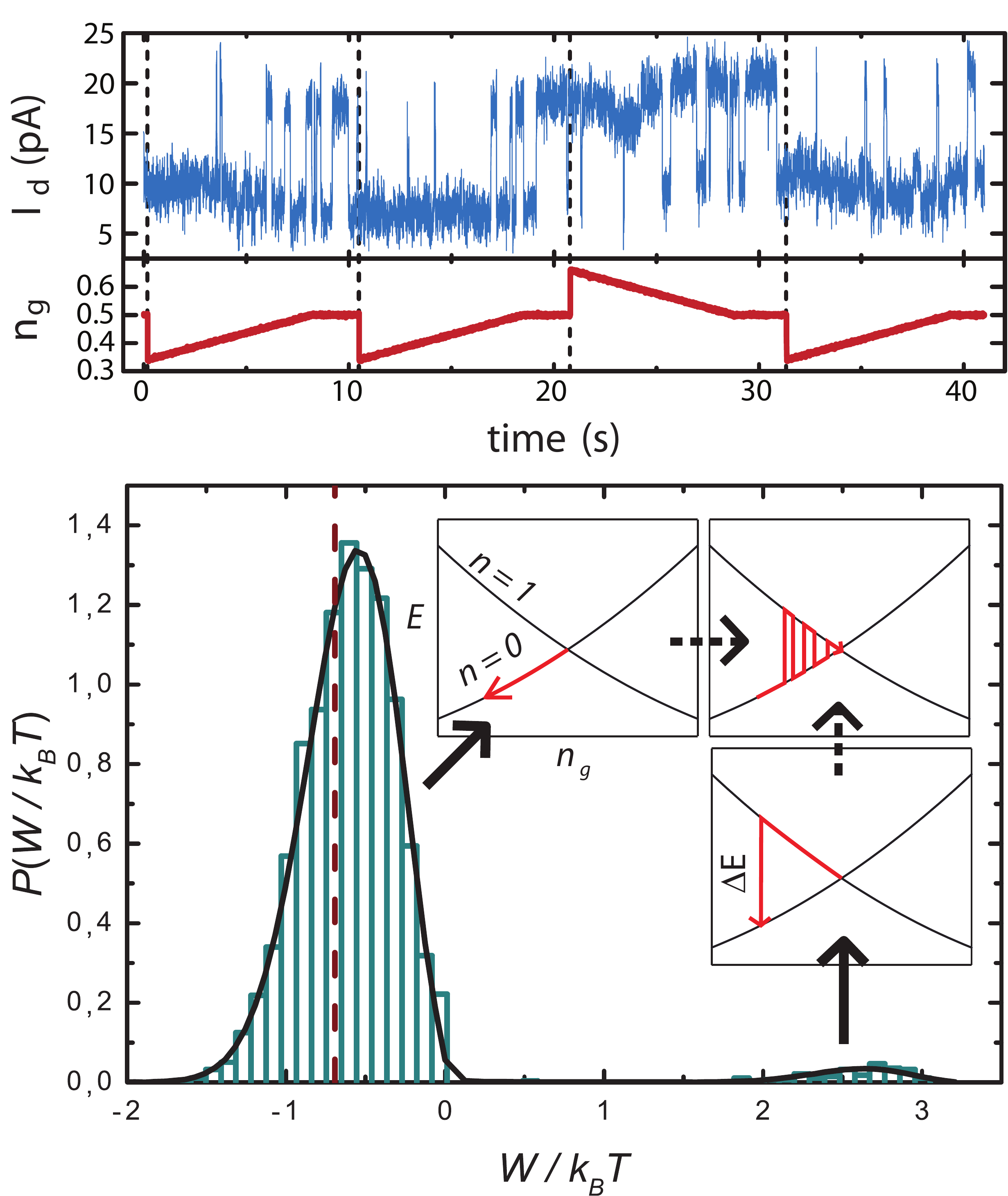}
  \caption{Top panel: A time trace of a Szilard's engine \cite{Szilard} with four realizations of the feedback control protocol. The vertical dashed lines mark the time instants for measuring the system state by observing the detector current $I_\text{d}$. Based on the measured $I_\text{d}$, the feedback drive rapidly changes $n_g$ to set the energy difference to $\Delta E \approx \pm 3.2 k_\text{B}T$, where the sign is determined by the measurement outcome.  Bottom panel: A histogram displaying the work done in about 3000 cycles of the Szilard's engine. The main peak at negative energies ($W \approx - k_BT\ln(2)$) is due to successful cycles whereas the weaker positive peak represents measurement and feedback errors. These errors counterbalance partly the successful ones leading eventually to about 75\% of the ideal work extraction of $k_BT\ln(2)$ on the average. Figure adapted from \cite{jvkPNAS}, Copyright 2014 National Academy of Sciences, U.S.A.}
 \label{fig:SEtraces}
\end{figure}


The drive protocol for the SEB considered here is the counterpart of erasure, namely using the one bit of information obtained from the electron position to extract $k_\text{B} T \ln(2)$ directly from the heat bath. Example traces of the protocol are shown in Fig. \ref{fig:SEtraces}. The cycle is similar to a Szilard's engine \cite{Szilard} by starting from the degeneracy point at thermal equilibrium, where an excess electron can reside either inside ($n$ = 1) or outside ($n$ = 0) of the SEB island with equal probability. The electron position is established from the detector current with threshold detection yielding a measurement outcome $m$. The feedback takes place as follows: if $m = 0$ was measured, $n_g$ is rapidly driven to introduce a positive energy difference $\Delta E = E_\text{fb}$ that effectively traps the electron to the $n = 0$ state. If $m = 1$ was measured, the energy difference is driven to be negative, $\Delta E = -E_\text{fb}$, trapping the electron to the $n = 1$ state. We refer to this first step as the 'fast drive'. Both protocols finish by slowly returning back to the degeneracy point, allowing thermal expansion of the electron. In the ideal limit of a perfect measurement and response, infinite $E_\text{fb}$, and infinitesimally slow drive, the work extracted in the cycle is $k_\text{B} T \ln (2)$ produced by the thermal excitations during the slow drive. In a practical experiment \cite{jvkPNAS}, we extract 75\% of this fundamental maximum.


For a practical experimental scheme, the measurement outcome $m$ has a strong but not perfect correlation to the actual charge state $n$. This is described by the probability $P(n = m) = 1 - \varepsilon$, where $\varepsilon$ is the error probability $P(n \neq m) = \varepsilon$, i.e. the probability to get an incorrect outcome. An important parameter that characterizes the maximum energy available is mutual information $I_\text{m}(n, m) = \ln(P(n | m)) - \ln(P(n))$. In a cyclic process with $\Delta F = 0$, the maximum work that can be extracted is $\langle -W \rangle \leq k_\text{B} T \langle I_\text{m}(n, m) \rangle$ \cite{sagawaueda}. Furthermore, irreversible feedback processes are described by Sagawa-Ueda equality \cite{sagawaueda}, also see chapter \chqfluct. The equality reads $\langle e^{-(W - \Delta F) / k_\text{B} T - I_\text{m}} \rangle = 1$. With a finite error probability, it is no longer optimal to drive with infinite $E_\text{fb}$. Every time the SEB is driven incorrectly, the fast drive excites the system leading to immediate relaxation and dissipation of $E_\text{fb}$. This is illustrated in the insets of Fig. \ref{fig:SEtraces} and explains the instances of positive work in the measured work distribution. The optimal $E_\text{fb}$ \cite{HorowitzParrondo} can be found by requiring that after the fast drive, the system is at thermal equilibrium. For example, if $n = 0$ was measured, the occupation probabilities are $P(0) = 1 - \varepsilon$ and $P(1) = \varepsilon$, which corresponds to the thermal equilibrium distribution with $\Delta E = k_\text{B} T \ln\left((1 - \varepsilon) / \varepsilon\right) \equiv E_\text{fb, opt}$ setting the optimal $E_\text{fb, opt}$.


In \cite{jvkPRLSE}, we consider a fixed $E_\text{fb}$ while the magnitude of error probability $\varepsilon$ is varied by changing the amount of detector signal averaging for the pre-feedback measurement. The error probability was estimated in a post-analysis, similar to the method of extracting the occupation probability as discussed in the third section of this chapter.. The extracted work $W$ increases linearly with decreasing error probability $\varepsilon$. However, the 'feedback efficiency' $\eta \equiv \langle -W \rangle / k_\text{B} T \langle I_\text{m}(n, m) \rangle$ exhibits a maximum in $\varepsilon$. This implies that the feedback process could be further improved by adjusting $E_\text{fb}$ accordingly. Furthermore, we find that for all $\varepsilon$, the Sagawa-Ueda relation holds.

\section{Single-electron box as an autonomous Maxwell's demon}

\begin{figure} [h!]
\includegraphics[width=12cm]{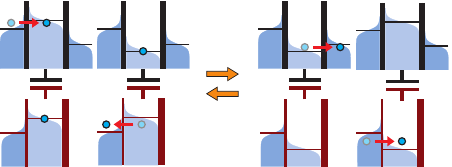}
  \caption{The operation principle of the autonomous Maxwell's demon. The top row shows the energy levels of the feedback-controlled SET, and the bottom row shows those for the SEB acting as a Maxwell's demon. The first and third steps show an electron tunneling in in the SET, which are processes that cost energy and therefore cool the system. The second and fourth step show an electron tunneling in the SEB, which are processes that release energy and heat up the demon. When the SET is voltage biased and the demon is designed to have a lower tunneling resistance, the above cycle is the most preferred one.}
 \label{fig:ADprinciple}
\end{figure}

An autonomous Maxwell's demon is a configuration where both the controlled system and the demon performing the measurements and the feedback are present \cite{Mandal, Barato, Strasberg}. In this section, we review the first experimental realization of an autonomous Maxwell's demon \cite{jvkAMD} by employing an SEB as the measurement and feedback control unit (the demon) for an SET. The demon is designed to apply a positive potential to the SET island when an electron is in there in order to trap it, and apply a negative potential to repel electrons from entering the island when an electron is not there. A small potential bias is applied to the SET to trigger charge transport, however due to the aforementioned feedback control by the demon, the SET cools down as all electron tunneling events cost energy, an effect that is verified by thermometry. A notable feature of this experiment on a Maxwell's demon is that its performance and heat flow is measured directly by observing a temperature change in the device. The operation principle is similar to the proposal in \cite{Strasberg} and is illustrated in Fig. \ref{fig:ADprinciple}. The two devices are gated such that the lowest energy configuration has a total of one electron in the whole system, i.e. an electron either in the SET or the SEB island, while excess negative charge (one electron in each island) and positive charge (zero electrons in each island) both have a higher energy equal to the mutual Coulomb interaction $J$ from Eq. \eqref{eq:coupled_devices}. This is achieved with $n_g = n_g^\text{det} = 0.5$, where $n_g$ now refers to the SET gate and $n_g^\text{det}$ to the demon gate. A similar configuration based on quantum dots has been used to realize an energy harvester \cite{Thierschmann} and a Maxwell's demon operating as a power generator \cite{Chida}.

We now consider the process cycle in Fig. \ref{fig:ADprinciple}, starting with one electron in the SEB and none in the SET, which is one of the degenerate ground states. The SET is voltage biased to trigger charge transport under condition $eV < J$ to ensure Coulomb blockade. An electron tunnels into the SET island from the source electrode with an energy cost $J/2 - eV/2$ that is provided by thermal excitations. The process costs energy, because the electron already present in the SEB island is repelling any electron that would enter the SET island. The SEB is designed to have a low tunneling resistance, i.e. a fast reaction time. This way, it can act as a Maxwell's demon with an electron tunneling out of the island in response to the transition in the SET. At this point, one electron is trapped in the SET and, similiarly to the initial setup, it can tunnel out to the drain electrode with an energy cost $J/2 - eV/2$. When it does, the demon reacts by an electron tunneling to the island, resuming back to the initial state and closing the operation cycle. During the cycle, the SET has been cooled down by an energy $J - eV$ due to the feedback-control of the demon. The thermodynamic cost of the cycle is the heat $J$ dissipated in the demon, restoring an agreement with Joule's law (total heat $Q = eV$) and the second law of thermodynamics.

\begin{figure} [h!]
\includegraphics[width=11cm]{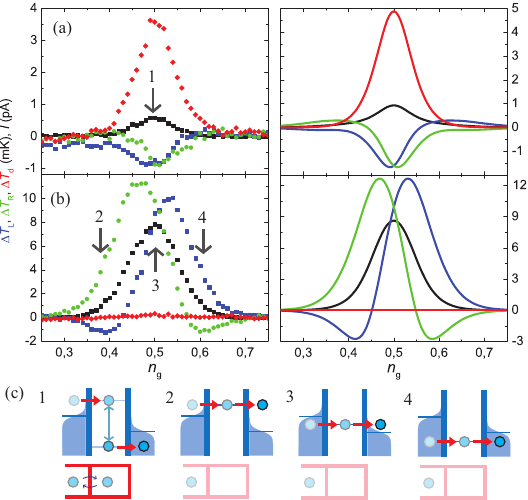}
  \caption{(a) - (b) Current through the SET (black), temperature of its left (blue) and right (green) electrode, and the temperature of the demon (red) as a function of the SET control parameter. Left panel shows measured data and the right shows calculated values. (a) shows the case where the SEB is allowed to operate as a demon, while in (b), the demon is deactivated by setting its control parameter to zero (the electron in the demon is trapped to the lead). (c) shows the schematics of the device operation at the settings indicated with arrows in (a) and (b). Reproduced figure with permission from \cite{jvkAMD}. Copyright 2015 by the American Physical Society.}
 \label{fig:ADresults}
\end{figure}


Note that the operation described above takes place internally in the circuit (autonomous operation), therefore there is no need to measure the system state with a charge detector. Furthermore, external feedback is not required, rather all external control parameters are constant. However, the device should be designed to have certain properties in order to optimize and observe the operation. First, the devices are designed to be fully normal metallic, introducing fast tunneling rates and enhanced cooling power to the device. The normal metallic junctions are realized by using the technique presented in \cite{jvkapl}. Second, the tunnel junction of the SEB is designed to have a lower tunneling resistance than the SET junctions to ensure fast feedback to the SET transitions. However, the resistances are designed to be higher than the quantum resistance $R_K = h / e^2 \approx 26$ k$\Omega$ to minimize the detrimental effects of co-tunneling \cite{Averin, Walldorf}. Third, the leads of both devices are thermally isolated while still permitting charge transport by interrupting them with superconducting aluminum leads \cite{Bardeen, Timofeev}. Finally, the temperatures of the system and the demon are readout with normal metal - insulator - superconductor thermometers \cite{Thermometry}. 


Figure \ref{fig:ADresults} shows the key results of the device operation. The data in Fig. \ref{fig:ADresults} (a) are obtained when the full feedback cycle described above takes place. We observe that even though heating would naively be expected from Joule's law as a finite charge current flows through the SET in the direction of voltage bias, both leads of the SET cools down as a result of the feedback control by the demon. The feedback control has a thermodynamic expense of heating the demon, apparent as a measured temperature rise. The total heat in the SET and the demon are in agreement with Joule's law.
For comparison, the data from a reference measurement with the demon deactivated ($n_g = 0$ for the demon) is shown in Fig. \ref{fig:ADresults} (b). While one-sided cooling in the SET is observed, the total heat generated in the system is always positive in the absence of feedback control. The temperature of the demon does not change, implying that there is no direct heat flow between the two devices. This concludes that in the feedback-controlled configuration (Fig. \ref{fig:ADresults} (a)) all the heat produced in the demon is a result of tunneling events and not of direct heat flow between the two devices. 

\bigskip

\acknowledgements

ACKNOWLEDGEMENTS

We acknowledge financial support from Academy of Finland contract number 312057.


\begin{thebibliography}{99}

\bibitem{rmp} F. Giazotto, T. T. Heikkil\"a, A. Luukanen, A. M. Savin, and J. P. Pekola, Rev. Mod. Phys. {\bf 78}, 217 (2006). 

\bibitem{wellstood} F. C. Wellstood, C. Urbina, and John Clarke,
Phys. Rev. B {\bf 49}, 5942 (1994).

\bibitem{pothier} H. Pothier, S. Gueron, Norman O. Birge, D. Esteve, and M. H. Devoret,
Phys. Rev. Lett. {\bf 79}, 3490 (1997).

\bibitem{jp} J. P. Pekola, Nat. Phys. {\bf 11}, 118 (2015).

\bibitem{njp} J. P. Pekola, P. Solinas, A. Shnirman, and D. V. Averin,
New J. Phys. {\bf 15}, 115006 (2013).

\bibitem{kung} B. Kung, C. R\"ossler, M. Beck, M. Marthaler, D. S. Golubev, Y. Utsumi, T. Ihn, and K. Ensslin, Phys. Rev. X {\bf 2}, 011001 (2012).

\bibitem{BCS} J. Bardeen, L. N. Cooper, and J. R. Schrieffer, Phys. Rev. \textbf{108}, 1175 (1957).

\bibitem{saira} O.-P. Saira, Y. Yoon, T. Tanttu, M. M\"ott\"onen, D. V. Averin and J. P. Pekola,
Phys. Rev. Lett. {\bf 109}, 180601 (2012).  

\bibitem{jvkapl} J. V. Koski, J. T. Peltonen, M. Meschke, and J. P. Pekola, Appl. Phys. Lett. \textbf{98}, 203501 (2011).

\bibitem{jarzynski} C. Jarzynski, Phys. Rev. Lett. \textbf{78}, 2690 (1997).

\bibitem{crooks} G. E. Crooks, Phys. Rev. E \textbf{60}, 2721 (1999).

\bibitem{Seifert} U. Seifert, Phys. Rev. Lett. \textbf{95}, 040602 (2005).

\bibitem{Schuler} S. Schuler, T. Speck, C. Tietz, J. Wrachtrup, and U. Seifert, Phys. Rev. Lett. \textbf{94}, 180602 (2005).

\bibitem{Tietz} C. Tietz, S. Schuler, T. Speck, U. Seifert, and J. Wrachtrup, Phys. Rev. Lett. \textbf{97}, 050602 (2006).

\bibitem{jvkNP} J. V. Koski, T. Sagawa, O-P. Saira, Y. Yoon, A. Kutvonen, P. Solinas, M. M\"ott\"onen, T. Ala-Nissila, and J. P. Pekola, Nat. Phys. \textbf{9}, 644 (2013).

\bibitem{Maisi} V. F. Maisi, S. V. Lotkhov, A. Kemppinen, A. Heimes, J. T. Muhonen, and J. P. Pekola, Phys. Rev. Lett. \textbf{111}, 147001 (2013).

\bibitem{sagawaueda} T. Sagawa and M. Ueda, Phys. Rev. Lett. \textbf{104}, 090602 (2010).

\bibitem{toyabe} S. Toyabe, T. Sagawa, M. Ueda, E. Muneyuki, and M. Sano, Nat. Phys. \textbf{6}, 988 (2010).

\bibitem{jvkPNAS} J. V. Koski, V. F. Maisi, J.P. Pekola and D. V. Averin, PNAS \textbf{111}, 13786 (2014).

\bibitem{Szilard} L. Szilard, Z. Phys. \textbf{53}, 840 (1929).

\bibitem{Landauer} R. Landauer, IBM J. Res. Develop. \textbf{5}, 183 (1961).

\bibitem{Berut} A. B\'erut, A. Arakelyan, A. Petrosyan, S. Ciliberto, R. Dillenschneider, and E. Lutz, Nature \textbf{483}, 187 (2012).

\bibitem{Jun} Y. Jun, M. Gavrilov, and J. Bechhoefer, Phys. Rev. Lett. \textbf{113}, 190601 (2014).

\bibitem{HorowitzParrondo} J. M. Horowitz and J. M. R. Parrondo, Euro Phys. Lett. \textbf{95}, 10005 (2011).

\bibitem{jvkPRLSE} J. V. Koski, V. F. Maisi, T. Sagawa, and J. P. Pekola, Phys. Rev. Lett. \textbf{113}, 030601 (2014).

\bibitem{Mandal} D. Mandal and C. Jarzynski, Proc. Nat. Acad. Sci. \textbf{109}, 11641 (2012).

\bibitem{Barato} A. C. Barato and U. Seifert, Europhys. Lett. \textbf{101}, 60001 (2013).

\bibitem{Strasberg} P. Strasberg, G. Schaller, T. Brandes, and M. Esposito, Phys. Rev. Lett. \textbf{110}, 040601 (2013).

\bibitem{jvkAMD} J. V. Koski, A. Kutvonen, I. M. Khaymovich, T. Ala-Nissila, and J. P. Pekola, Phys. Rev. Lett. \textbf{115}, 260602 (2015).

\bibitem{Thierschmann} H. Thierschmann, R. S\'anchez, B. Sothmann, F. Arnold, C. Heyn, W. Hansen, H. Buhmann, and L. W. Molenkamp, Nat. Nano. \textbf{10}, 854 (2015).

\bibitem{Chida} K. Chida, S. Desai, K. Nishiguchi, and A. Fujiwara, Nat. Comm. \textbf{8}, 15310 (2017).

\bibitem{Averin} D. V. Averin and Y. V. Nazarov, Phys. Rev. Lett. \textbf{65}, 2446 (1990).

\bibitem{Walldorf} N. Walldorf, A.-P. Jauho, and K. Kaasbjerg, Phys. Rev. B \textbf{96}, 115415 (2017).

\bibitem{Bardeen} J. Bardeen, G. Rickayzen, and L. Tewordt, Phys. Rev. \textbf{113}, 982 (1959).

\bibitem{Timofeev} A. V. Timofeev, M. Helle, M. Meschke, M. M\"ott\"onen, and J. P. Pekola, Phys. Rev. Lett.\textbf{102}, 200801 (2009)

\bibitem{Thermometry} M. Nahum, T. M. Eiles, and J. M. Martinis, Appl. Phys. Lett. \textbf{65}, 3123 (1994).

\bibitem{horowitz} J. M. Horowitz and M. Esposito, Phys. Rev. X \textbf{4}, 031015 (2014).

\bibitem{kutvonen} A. Kutvonen, J. V. Koski, and T. Ala-Nissila, Sci. Rep. \textbf{6}, 21126 (2016).

\end{thebibliography}
\end{document}